\documentclass[
 reprint,
 floatfix,
 amsmath,amssymb,
 aps,
 prd,
english
]{revtex4-2}

\usepackage{graphicx}
\usepackage{dcolumn}
\usepackage{bm}
\usepackage{csquotes}
\usepackage{verbatim}
\usepackage{hyperref}

\newcommand{\sun}{\odot}

\begin{document}

\title{FORGE'D IN THE EARLY UNIVERSE: THE EFFECT OF PROTOSTELLAR OUTFLOWS ON POP III ACCRETION}0

\author{Yasmine J. Meziani}
\email{ymeziani@caltech.edu}
\affiliation{TAPIR, California Institute of Technology, Mailcode 350-17, Pasadena, CA 91125, USA}

\author{Philip F. Hopkins}
\affiliation{TAPIR, California Institute of Technology, Mailcode 350-17, Pasadena, CA 91125, USA}

\author{Michael Y. Grudić}
\affiliation{Center for Computational Astrophysics, Flatiron Institute, 162 5th Ave, New York, NY 10010, USA}

\author{Claude-André Faucher-Giguère}
\affiliation{CIERA and Department of Physics and Astronomy, Northwestern University, Evanston, IL 60201, USA}

\author{Shivan Khullar}
\affiliation{Center for Computational Astrophysics, Flatiron Institute, 162 5th Ave, New York, NY 10010, USA}

\author{Pratik J. Gandhi}
\affiliation{Department of Astronomy, Yale University, New Haven, CT 06520, USA}

\date{\today}


\begin{abstract}
We present a cosmological zoom-in radiation magneto-hydrodynamic (RMHD) simulation, using FORGE'd in FIRE, that follows the formation, growth, and evolution of a single metal-free Pop.~III (proto)star at redshift $z \sim 14$. The simulation captures a rotationally supported circumstellar disk and protostellar jets, both resolved down to $<100\;\mathrm{au}$ scales. We find the star grows to $\sim 27 \; \mathrm{M}_{\odot}$ over $31,000 \; \mathrm{years}$, with its final mass regulated by accretion and protostellar jets. Protostellar jets form because the magnetic mass-to-flux ratio lies within the regime that allows jet launching, and they are further enabled by a rotating circumstellar disk with sufficient gas–magnetic-field coupling, both present in this simulation. These jets regulate accretion onto the (proto)star and drive outflows that collide with infalling gas, slowing inflow at large radii due to the substantial momentum they carry. A circumstellar disk forms, extending out to $\sim 0.01 \; \mathrm{pc}$, which remains gravitationally stable (Q $>>1$). The stability of the disk is maintained through both thermal support and turbulence. In this paper we focus on how jets play a critical role not only in shaping the final masses of Pop.~III stars but also in directly influencing their surroundings by regulating accretion. These results will provide important insights into the initial mass function and feedback processes in the earliest star-forming regions of the Universe.
\end{abstract}

\maketitle

\section{Introduction} \label{sec:Intro}
The formation of the first generation of stars, known as Population III (Pop.~III) stars, represents a critical frontier in our understanding of early cosmic structure formation. Forming from metal-free gas at high redshifts ($z \sim 10-20$), these stars set the initial conditions for subsequent galaxy formation and chemical enrichment \citep{ostriker_reheating_1996, gnedin_reionization_1997, ferrara_mixing_2000, madau_early_2001, mori_early_2002, bromm_formation_2003, furlanetto_metal_2003, mackey_three_2003, scannapieco_detectability_2003, wada_feedback_2003, bromm_first_2004, yoshida_era_2004, beers_discovery_2005, wise2008resolving, clark_formation_2011, stacy_constraining_2013}. Despite their central role, the formation and evolution of Pop.~III stars remain poorly constrained because direct observations are unlikely with current facilities \citep{schauer2020ultimately, katz_challenges_2023, trussler_observability_2023, katz_closing_2025, glover_first_2026}, and theoretical models are limited by the enormous dynamic range in time and space required to connect cosmological structure formation to (proto)stellar physics \citep{bromm_first_2011, greif_numerical_2015, glover_first_2026}.

Recent advances in large-scale galaxy formation simulations, such as FIRE \citep{hopkins_galaxies_2014, hopkins_fire-2_2018, hopkins_fire-3_2023}, and small-scale star formation simulations, such as STARFORGE \citep{grudic_starforge_2021}, have begun to close this gap. Recent simulations build on these advancements---the FORGE’d in FIRE simulations enable zoom-in simulations of individual stars within a cosmological context \citep{hopkins2023forgeI, hopkins2023forgeII, hopkins2024forgeIII}. This approach captures both the large-scale assembly of primordial gas in dark matter minihalos and the detailed (proto)stellar physics governing accretion, disk formation, and outflows (jets and winds). Additionally, simulations using STARFORGE are able to produce realistic initial mass functions (IMF) and multiplicities in metal-rich analogues of nearby star-forming clouds \citep{guszejnov_starforge_2021, guszejnov_effects_2022}. Understanding the IMF of Pop.~III stars is vital for determining their contribution to the earliest stages of cosmic reionization 
and for dictating the chemical enrichment patterns established by the first supernovae \citep{ostriker_reheating_1996, gnedin_reionization_1997, mackey_three_2003, bromm_first_2004, bromm_first_2011}.

Many previous studies, employing various physical models, have argued that factors such as magnetic fields, radiation, and other processes play important roles \citep[e.g.,][]{machida_first_2006, machida_formation_2013, stacy_building_2016, sharda_importance_2020, sharda_magnetic_2021, sharda_impact_2022,
klessen_first_2023, sharda_population_2025}. However, these studies often involve compromises in several areas: the initial and boundary conditions (which may be idealized as a simple collapsing cloud), limited physics (such as omitting radiation, magnetic fields, jets, or using simplified cooling), and timescales (no main sequence evolution). Importantly, almost all previous work used a combination of physics and numerical methods known not to produce realistic observed IMFs or stellar multiplicity for stars in nearby metal-rich star-forming regions where observations are plentiful. The new simulation presented in this paper extrapolates the predictive power of STARFORGE simulations to unobserved Pop.~III systems while including a realistic cosmological environment simulated with FIRE physics on larger scales.

While many previous studies of Pop.~III star formation have aimed to predict the initial mass function, these efforts have generally not demonstrated converged IMF predictions across varying environments. Accurate IMF prediction has only recently become possible at $z=0$, and only if one includes all the relevant physical processes listed above. Additionally, there has been debate over whether Pop.~III stars exhibit protostellar jets. Previous simulations that did include jets often treated them as simple mass-removal prescriptions, with the physical jet itself remaining unresolved.

In this paper, we present the first application of FORGE’d in FIRE to the formation of an individually resolved Pop.~III (proto)star around $z \sim 14$, within a fully cosmological zoom-in setup. These simulations include, but are not limited to, protostellar jets, radiation, magnetic fields, and thermo-chemistry. Here, we focus on the first star, highlighting its circumstellar disk and outflow structures, accretion history, and stellar growth onto the main sequence, to understand some of the key physics above. Future work will explore the predicted IMF, stellar multiplicity, and eventual supernovae of a group of Pop.~III stars.

The structure of this paper is as follows. In Section \ref{sec:Simulations}, we describe the simulation framework and the physics included. Section \ref{sec:Results} presents a case study of a Pop.~III (proto)star, detailing its stellar properties, associated feedback, and circumstellar disk. We discuss the conclusions of our results and directions for future work in Section \ref{sec:Conclusions}.

\section{Simulations} \label{sec:Simulations}

\begin{figure*}[t!]  
    \centering
    \includegraphics[width=\textwidth]{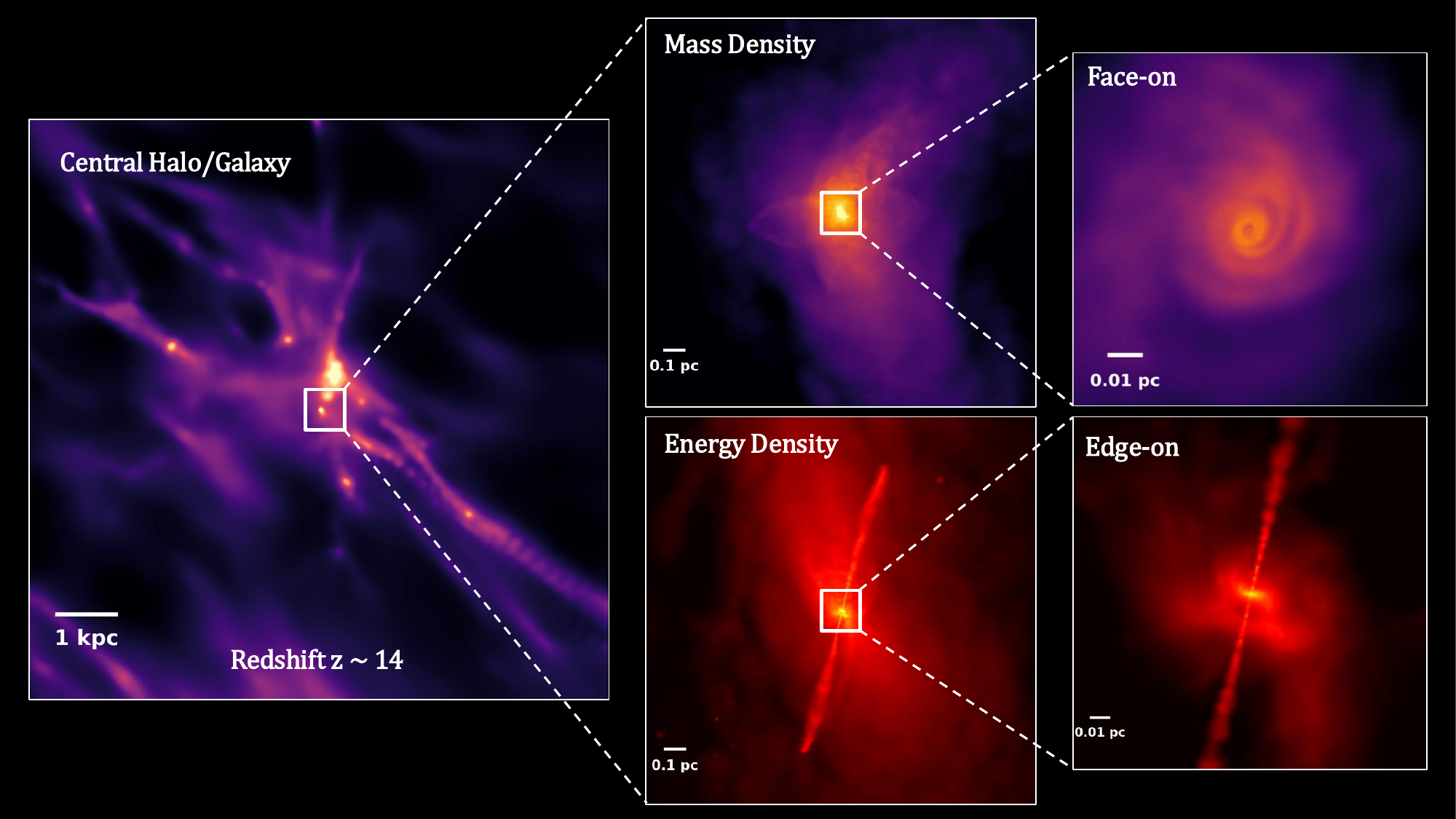}
    \caption{Visualizations of the gas mass and gas kinetic energy surface densities in our cosmological simulation at redshift $z \sim 14$. We highlight the dynamic range resolved in our simulation by zooming in from galaxy-wide scales on the left to the scales of the circumstellar disk on the right. In both density plots, the color increases from black to white (different properties use different color schemes) on a logarithmic scale, where each panel uses its own dynamic range. For the mass density plots (differentiated with a black through white color scheme) the median gas surface density ($\Sigma_{\mathrm{median}}$) in the top left, top middle, and top right panels are $1.5\times10^{-1}\, M_{\odot}\,\mathrm{pc}^{-2}$, $1.7\times10^{3} \, M_{\odot}\,\mathrm{pc}^{-2}$, and $2.3\times10^{4}\, M_{\odot}\,\mathrm{pc}^{-2}$ respectively. For the bottom middle and bottom right panels, we see the bipolar outflows from protostellar jets (in bright red) extend out to $\sim 1 \,\mathrm{pc}$. The top right panel highlights a circumstellar disk extending out to $\sim 0.01 \, \mathrm{pc}.$}
    \label{fig:visualization}
\end{figure*}

The cosmological simulations follow FORGE’d in FIRE \citep{hopkins2023forgeI, hopkins2023forgeII, hopkins2024forgeIII}. On large scales, the simulations adopt the FIRE-3 physics model \citep{hopkins_fire-3_2023} within the Feedback In Realistic Environments framework (FIRE; \citealt{hopkins_galaxies_2014, hopkins_fire-2_2018, hopkins_fire-3_2023}), using the m09 initial conditions from earlier FIRE studies \citep{hopkins_galaxies_2014}. In the refined regions, the full-physics resolution follows the Star Formation in Gaseous Environments (STARFORGE) project \citep{grudic_starforge_2021}. \citet{grudic_starforge_2021, hopkins_fire-3_2023, hopkins2023forgeI} present all numerical method details, but we briefly review the most relevant aspects here.

We run the simulations in this paper with \texttt{GIZMO} \citep{hopkins_new_2015, hopkins_new_2017}. The code evolves the coupled equations of gravity, magneto-hydrodynamics, and radiation transport, using the meshless finite-mass (MFM) method \citep{hopkins_accurate_2016, hopkins_anisotropic_2017}. For gravity, \texttt{GIZMO} employs adaptive Lagrangian force softening that matches the hydrodynamic and force resolution for gas cells, uses fixed softenings for collisionless particles, and integrates the equations of motion with a fifth-order Hermite scheme to accurately capture close gravitational encounters, such as binaries \citep{grudic_starforge_2021}.
We initialize the simulations with trace seed cosmological magnetic fields of $\sim 10^{-15}\, \mathrm{G}$, which amplify self-consistently in the ISM, using the constrained-gradient MHD solver in \citet{hopkins_constrained-gradient_2016}. 
The code follows radiation transport with an M1 moments method across multiple frequency bands (photo-ionizing, Lyman-Werner, photo-electric, NUV, optical \& near-IR, and adaptive [multi-wavelength grey-body] FIR), enabling self-consistent coupling between gas, dust, and radiation.

The code fully couples radiation to both dust and gas with a fixed dust-to-metals ratio. FORGE'd in FIRE \citep{hopkins2023forgeI, hopkins2023forgeII, hopkins2024forgeIII} has extended 
thermo-chemistry and opacities 
to temperatures of $T \sim 1$--$10^{10}$~K, densities $n \sim 10^{-8}$--$10^{24}$~cm$^{-3}$, and metallicities $Z=0-10 \; \mathrm{Z}_{\odot}$. Thermo-chemistry includes a simplified radiation network that accounts for dust, atomic and fine-structure cooling, and the abundance evolution of fully and partially ionized and atomic H and He, as well as HD (see, e.g., \citealt{john_continuous_1988}; \citealt{glover_star_2007, glover_uncertainties_2008}; and references in \citealt{hopkins_fire-3_2023}). H$_2$ cooling and chemistry is evolved on-the-fly using the treatment adopted in STARFORGE \citep{grudic_starforge_2021}. 

Sink particles (see \citealt{grudic_starforge_2021} for a full description) form to represent individual (proto)stars which follow their own (proto)stellar evolution tracks as they grow and accrete, proceed through the main sequence, and end their main sequence lives as stellar remnants or SNe. Gas cells become eligible for (proto)star formation if they meet the criteria described by \citet{grudic_starforge_2021}: high-density, self-gravitating, collapsing, isolated from other sinks, experiencing compressive tidal forces in all directions, and capable of collapsing faster than the local orbital or accretion timescales. For accretion, a gas cell must satisfy angular momentum and boundness criteria and lie within the sink radius of the star. Accreted gas first enters 
a \enquote{reservoir} which can accrete onto the protostar directly or be ejected by jets, where the \citet{shu_self-similar_1977} accretion timescale sets the depletion time.

Once formed, sink particles include a wide range of processes essential for protostellar evolution, including accretion-powered collimated bipolar outflows; radiatively driven stellar winds on the main sequence; the evolution of radiation in all bands;
and supernova explosions for massive stars at end of life. \citet{grudic_starforge_2021} and \citet{guszejnov_effects_2022} present detailed physical and numerical prescriptions for jets, winds, radiation, and supernovae; for completeness, we briefly summarize the relevant physics below. 


Protostellar jet modeling follows \citet{cunningham_radiation-hydrodynamic_2011}, motivated by high-resolution simulations and observations of individual jets. The model diverts a fraction of mass accreted by the envelope-disc-star system to the jet at a fixed value of $f_{w} = 0.3$. The jet launch velocity, $v_{jet} = f_{K} (GM_{\star}/R_{\star})^{1/2}$, uses a fixed fraction $f_{K} = 0.3$ of the Keplerian velocity at the protostellar radius $R_{\star}$. The jets have a narrow collimation along the angular momentum axis of the sink, with a typical opening angle $\theta_0 = 0.01$ \citep{matzner_bipolar_1999}.

There is ongoing debate about whether Pop.~III stars are able to launch their own jets, as this depends on whether some trace magnetic field exists for amplification
in this stage. In the following sections, we show that the magnetic mass-to-flux ratios around the (proto)star fall well within the range needed for strong jet launching, indicating that, at least for the simple B-field model here, we expect jet launching and include jets. 

We inject stellar winds launched by main-sequence stars more massive than $2 \, M_{\odot}$, following a phenomenological prescription for the wind mass-loss rate
(see methods paper [\citealt{grudic_starforge_2021}] for details), where $\dot{M}_{wind} \propto Z_{Fe}^{0.7}$ in terms of the stellar Fe abundance; these winds are therefore nil for the models considered here (i.e, we neglect winds driven by e.g., Ly$\alpha$ radiation pressure). For supernovae, the model assumes that all stars more massive than $8 \, M_{\odot}$ undergo a $10^{51} \; \mathrm{erg}$ explosion at the end of their lifetimes; however, no supernovae occur in the simulation we present in this paper since the star only evolves for $\sim 30,000 \; \rm years$. 

\subsection{Initial Conditions and Refinement Choices} \label{subsec:IC and Refinement}

\begin{figure}[ht!] 
\includegraphics[width=\columnwidth]{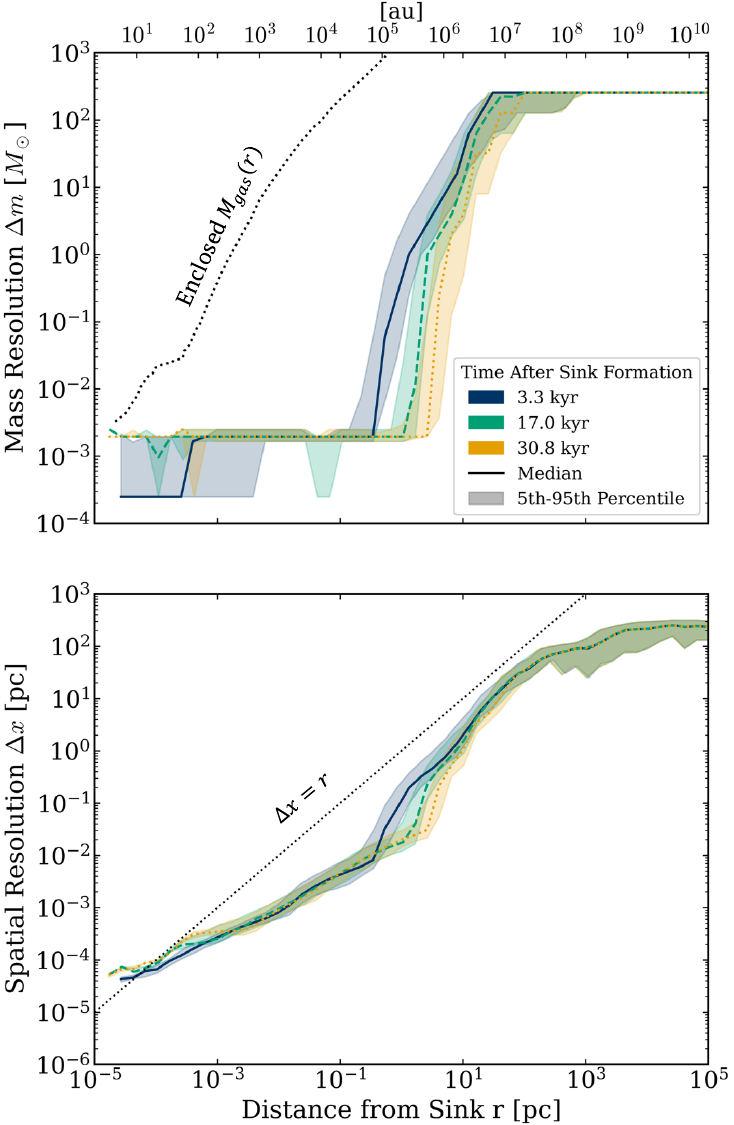}
\caption{Effective resolution of the simulation as a function of radial distance from the sink (r). We show the median and $90\%$ inclusion interval (\textit{shaded}) of the mass resolution and spatial resolution of the gas at various times after sink formation. The top panel plots the mass resolution ($\Delta m$), defined as the gas cell mass. We also include a reference line showing the total enclosed gas mass within a sphere of radius $r$ centered on the sink (M$_{\mathrm{gas}}(r)$). We define the spatial resolution (bottom panel) as $\Delta x \equiv (\Delta m /\rho)^{1/3}$. The resolution rapidly refines as the distance from the sink decreases. The regions near the (proto)star and its disk reach a mass resolution of $\Delta m < 3 \times 10^{-3} \; M_{\odot}$.
}
\label{fig:resolution}
\end{figure}
Our initial condition is a fully cosmological \enquote{zoom-in} simulation, evolving a large box of length $5$ comoving Mpc from redshifts $z \gtrsim 100$,  and concentrating resolution within an irregular Lagrangian volume centered on a target halo. We specifically center around halo \enquote{\textbf{m09}}, an ultra-faint dwarf galaxy at $z=0$ that we extracted from the FIRE cosmological simulation suite \citep{hopkins_galaxies_2014}. The galaxy has a total halo mass of $M_{\rm halo} \sim 2 \times 10^{9} \; M_{\odot}$ at $z=0$, and is simulated with an initial ($z=0$) baryonic particle mass of $m_{\rm b} = 1.3\times10^2 \; M_{\odot}$ and dark matter particle mass of $m_{\rm dm} = 6.34 \times 10^2 \; M_{\odot}$. At the redshift when the first star forms ($z \sim 14$), the host halo mass is on the order of $10^{6} \; \rm M_{\odot}$. We choose this well-studied galaxy \citep{hopkins_galaxies_2014, wheeler_sweating_2015, wheeler_be_2019, shen_dissipative_2021, gandhi2022exploring} because it is an ultra-faint dwarf; such galaxies are relics of the early universe owing to their uniformly metal-poor, ancient stellar populations, ideal for studying the formation of the first stars \citep{ricotti_formation_2005, bovill_pre-reionization_2009, okamoto_stellar_2012, brown_quenching_2014, weisz_star_2014, frebel2015near, weisz_local_2017, durbin2025hst}. Figure \ref{fig:visualization} shows the cosmological setting and small-scale dynamics. 

The refinement techniques build on the “hyper-refinement” framework developed in previous FORGE'd in FIRE studies of galaxy nuclei \citep[e.g.,][]{angles-alcazar_cosmological_2021, franchini_resolving_2022, hopkins2023forgeI, hopkins2023forgeII, hopkins2024forgeIII}. The simulation starts at $z \sim 100$ and evolves to $z \sim 14$. The refinement adaptively adjusts the gas particle mass resolution based on local physical conditions. In particular, we refine particles when their mass exceeds a multiple of $0.001$ of the local Jeans mass, ensuring that Jeans-unstable gas is always resolved with $\sim 1000 \; \mathrm{cells}$.
Additionally, the refinement increases with decreasing distance from the nearest sink particle, so regions surrounding collapsing objects maintain high resolution and do not de-refine even if the density decreases or temperature increases. In particular, any gas $< 10 \mathrm{pc}$ of the sink is resolved at the STARFORGE limit \citep{hopkins2023forgeI}. 

Figure \ref{fig:resolution} shows the effective mass and spatial resolution of the simulations as a function of sink-centric radius r, evaluated at $3.3$, $17.0$, and $30.8$ kyr after sink formation. For typical gas cells in our simulations around the (proto)star, the mass resolution is $\Delta m_{\rm gas} \sim 2 \times 10^{-3}\,\mathrm{M}_{\odot}$. The refinement scheme further resolves cells associated with protostellar jets and stellar winds to a mass resolution of $\sim10^{-4} \, \mathrm{M}_{\odot}$ \citep{grudic_starforge_2021}. The spatial resolution in the vicinity of the star is on the order of $10^{-4}\; \mathrm{pc}$. 

\section{Results} \label{sec:Results}

Figure \ref{fig:visualization} shows the projected gas and energy densities at different scales, highlighting the circumstellar disk and jets. We see that the jets extend out to $\sim 1\,\mathrm{pc}$, while the disk extends out to $\sim 0.01\, \mathrm{pc}$. Figure \ref{fig:resolution} shows the mass and spatial resolution as a function of radial distance from the sink. The cosmological dynamics span low-resolution regions far from the (proto)star, at $\gtrsim 10 \; \mathrm{pc}$, while the high-resolution, explicitly evolved domain spans $\sim 10^{-4.5} - 0.1 \; \mathrm{pc}$ (a few to $10^4 \; \mathrm{au}$). 

\subsection{Stellar Properties} \label{subsec:Stellar Properties}
\begin{figure*}[t!]
    \centering
    \includegraphics[width=\textwidth]{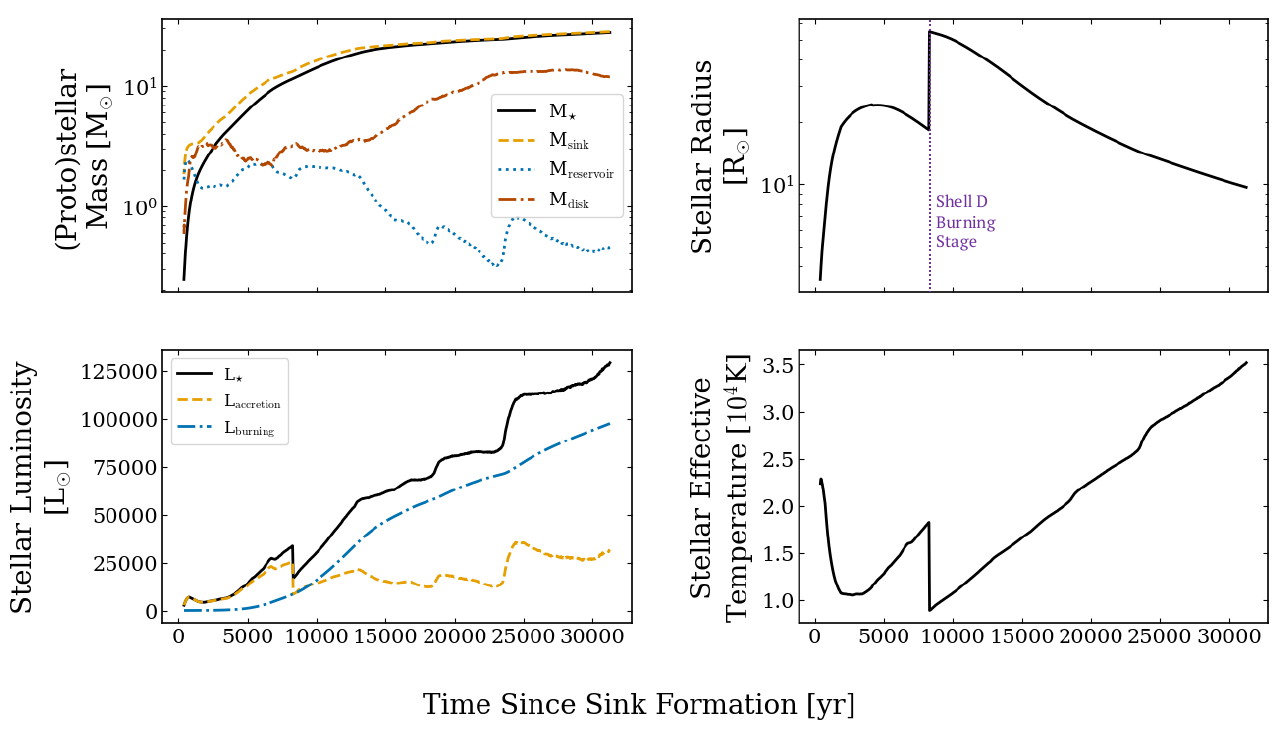}
    \caption{Evolution of the sink particle's physical properties over time. 
    \textit{Top Left:} The plot shows the evolution of the sink mass, the reservoir mass, the stellar mass, and the circumstellar disk mass. Here, the sink particle represents a subgrid object that accretes gas from its surroundings, with the sink mass defined as the sum of the stellar mass and the reservoir mass. The reservoir is the mass of gas accreted onto the sink that is either incorporated into the star or diverted to a \enquote{jet reservoir} used to launch jets. The stellar mass corresponds to the mass accreted onto the star from the reservoir at a rate of $(1 - f_W)\dot{M}_{\mathrm{acc}}$, where $f_W = 0.3$ and $\dot{M}_{\mathrm{acc}}$ is the accretion rate \citep{grudic_starforge_2021}. The disk mass represents the sum of gas particles traveling on approximately circular orbits around the sink. The total stellar mass grows steadily, reaching a final mass of $\sim 27\,M_{\odot}$. \textit{Bottom Left:} We plot the total stellar luminosity L$_{\star}$, the accretion luminosity L$_{\mathrm{accretion}} = GM_{\star} \dot{M}/R_{\star}$, and the burning luminosity L$_{\mathrm{burning}} = \mathrm{L}_{\star} - \mathrm{L}_{\mathrm{accretion}}$. The total luminosity increases to $\sim$1.3 $\times 10^{5}$\,L$_{\odot}$ by the end of the simulation. The accretion luminosity dominates during the first $9000$ years after sink formation, after which the burning luminosity becomes the primary contribution. \textit{Top Right:} The stellar radius expands rapidly for the first $9000 \; \mathrm{years}$ due to accretion from the surrounding envelope, and then begins to contract. \textit{Bottom Right:} Stellar effective (photospheric) temperature (defined such that $\mathrm{L}_{\star} = 4\pi R_{\star}^2 \sigma_{B} T_{eff}^{4}$).
    \label{fig:sink properties}
    }
\end{figure*}

We study the growth of one Pop.~III (proto)star, which is the first and only star to form in the simulation by the time it is stopped. At the times analyzed in this paper, the star remains in its pre-main-sequence phase. Specifically, at the last time analyzed here, the (proto)star is in the Shell D burning stage, following the (proto)stellar evolution sequence described in \citet{offner_effects_2009}. While the validity of any protostellar evolution model for Pop.~III stars is inherently uncertain, this is a general limitation of such models rather than a flaw in the present approach. Another commonly used protostellar evolution model for Pop.~III stars is that of \citet{haemmerle_rotation_2018}, which focuses on supermassive Pop.~III stars. To estimate the approximate uncertainty in our results, one could compare the stellar radii predicted by the model in \citet{grudic_starforge_2021} with those predicted by \citet{haemmerle_rotation_2018} for stars of similar mass.

\begin{figure*}[t!]
  \centering
  \includegraphics[width=\textwidth]{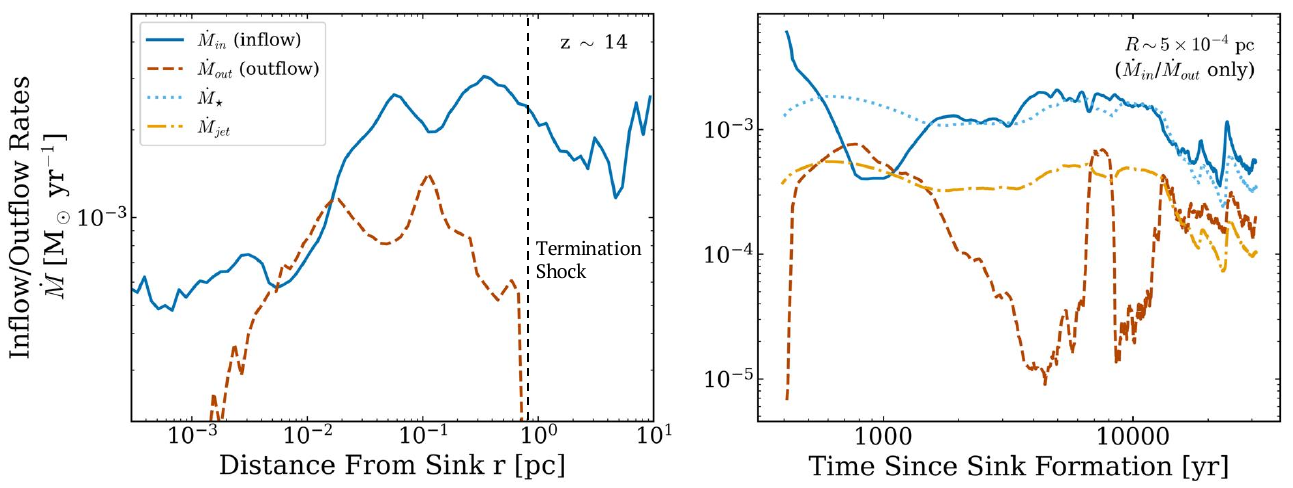}
  \caption{Mass flow rate (accretion rate) $\dot{M}$, showing the total inflow rate $\dot{M}_{\mathrm{in}}$ (blue) and total outflow rate $\dot{M}_{\mathrm{out}}$ (brown). \textit{Left:} The left panel shows inflow and outflow rates that we measure in spherical shells at increasing distances from the sink. Outflow slightly dominates between $5\times 10^{-3}$ and $5\times 10^{-2}\,\mathrm{pc}$, beyond which inflow becomes dominant. This indicates that jets efficiently clear gas near the sink, allowing inflow to dominate farther out. The sharp drop in outward accretion near $1\,\mathrm{pc}$ corresponds to the termination shock visible in Figure \ref{fig:radial velocity}, where the gas stalls and no longer propagates outward. \textit{Right:} The right panel shows the time evolution of inflow and outflow rates following sink formation at a fixed radial distance of $\sim115\; \mathrm{au}$ from the sink. It also includes the stellar accretion rate, $\dot{M}_{\star}$, and the outward jet accretion rate, $\dot{M}_{jet}$. The inflow accretion rate varies between $\sim 3 \times 10^{-4}$ and $\sim 5 \times 10^{-3}\,M_{\odot}\,\mathrm{yr}^{-1}$ during the first $\sim 1000\,\mathrm{years}$ after formation. During this same early period, the onset of protostellar jets drives a sharp rise in the outflow rate, reaching up to $\sim 10^{-3}\,M_{\odot}\,\mathrm{yr}^{-1}$. Thereafter, the outflow rate generally ranges between $10^{-5}$ and $10^{-3}\,M_{\odot}\,\mathrm{yr}^{-1}$.}
  \label{fig:accretion}
\end{figure*}

The (proto)star initially has an inward accretion rate of $\dot{M}_\mathrm{in} \sim 5 \times 10^{-3} \, M_{\odot} \, \mathrm{yr}^{-1}$, which decreases to $\sim 5 \times 10^{-4} \, M_{\odot} \, \mathrm{yr}^{-1}$ within the first $1000 \; \mathrm{years}$ (see the right panel of Figure~\ref{fig:accretion}). This early decline in accretion reflects the depletion of the immediate gas reservoir surrounding the (proto)star (Section~\ref{subsec:Circumstellar Disk}). Around $1000 \; \mathrm{years}$ after sink formation, the inward accretion rate rises again to $\sim 10^{-3} \, M_{\odot} \, \mathrm{yr}^{-1}$, possibly driven by infall from the outer disk or envelope. By approximately $31,000 \, \mathrm{years}$, the inward accretion rate gradually declines to $\sim 5 \times 10^{-4} \, M_{\odot} \, \mathrm{yr}^{-1}$, consistent with the (proto)star approaching a more stable mass.

The (proto)star exhibits rapid early growth, reaching a mass of $\sim 22 \; M_{\odot}$ within its first $17,000 \; \mathrm{years}$ of evolution. Its growth rate then slows, gradually increasing to $\sim 27 \, M_{\odot}$ by $30,000 \; \mathrm{years}$. As shown in Figure \ref{fig:sink properties}, the stellar mass begins to plateau around $30,000 \; \mathrm{years}$ after formation. Although further growth is expected, substantially larger final masses would require accretion to remain effective over $\gtrsim 10^{4} \; \rm years$. At the latest time, the Kelvin-Helmholtz timescale t$_{\rm KH} \sim 1.8 \times 10^4 \; \rm years$, is shorter than the accretion timescale t$_{\rm acc} \sim 5.4 \times 10^4 \; \rm years$, indicating that the (proto)star can thermally contract faster than it gains mass. Since t$_{\rm KH} < \rm t_{acc}$ at the end of the simulation, the (proto)star will become a main sequence star before it can further accrete an appreciable amount of mass.

From Figure \ref{fig:sink properties} we can see that the initial luminosity is primarily from accretion (L$_{\mathrm{accretion}}$). The transition from accretion-dominated luminosity to burning-dominated luminosity occurs at $\sim 9000 \; \mathrm{years}$ with the start of the \enquote{shell D burning stage} \citep{offner_effects_2009}. Deuterium burning causes a rapid increase in the stellar radius (Figure \ref{fig:sink properties}, top right). Subsequently, the luminosity from nuclear burning (L$_{\mathrm{burning}}$) exceeds the accretion luminosity, becoming the dominant component of the (proto)star's total energy output. 

Initially, infrared (IR) emission dominates the (proto)stellar radiation. In metal-free, dust-free gas, opacities are extremely small and the gas remains optically thin ($\tau << 1$) to its dominant cooling radiation. As a result, IR photons escape efficiently without significant energy deposition. Following deuterium shell burning, the (proto)star exhibits an initial drop followed by a steady increase in temperature (Figure \ref{fig:sink properties}, bottom right). This gradual temperature increase shifts its spectral energy distribution to shorter wavelengths, producing a high flux of ionizing and Lyman-Werner (LW) photons.

In contrast to IR, both ionizing and LW radiation have significant opacity in metal-free gas. The absorption of ionizing photons increases thermal pressure, while LW photons dissociate H$_{2}$. This dissociation destroys the primary cooling mechanism and prevents further gas collapse. This combination of radiative feedback and protostellar jets disrupts accretion onto the star and thus regulates the (proto)star's final stellar mass.

At a stellar mass of $\sim 27 \, M_{\odot}$, the star has a radius of $\sim 10\, R_{\odot}$, an effective temperature of $\sim 3.5 \times 10^{4} \, \mathrm{K}$, and a luminosity of $\sim 1.3 \times 10^{5} \, L_{\odot}$. The combination of these properties with the reduced accretion rate indicates that the star has entered a phase of self-regulated growth, where feedback from jets limits further mass growth rather than through depletion of the surrounding disk. The inferred final mass is broadly consistent with previous Pop.~III studies, which typically find main-sequence masses in the $0.1 - 50 \, M_{\odot}$ mass range \citep[e.g.,][]{stacy_building_2016, latif_birth_2022, sharda_population_2025, sharda_magnetic_2025}. In contrast, early Pop.~III simulations predicted characteristic stellar masses of $\sim 10^2$–$10^3\,M_{\odot}$ \citep[e.g.,][]{bromm_forming_1999, nakamura_initial_2001, abel_formation_2002, bromm_first_2004}. 

\subsection{Inflow Dynamics} \label{subsec:Circumstellar Disk}

\begin{figure*}[h] 
\includegraphics[width=\textwidth]{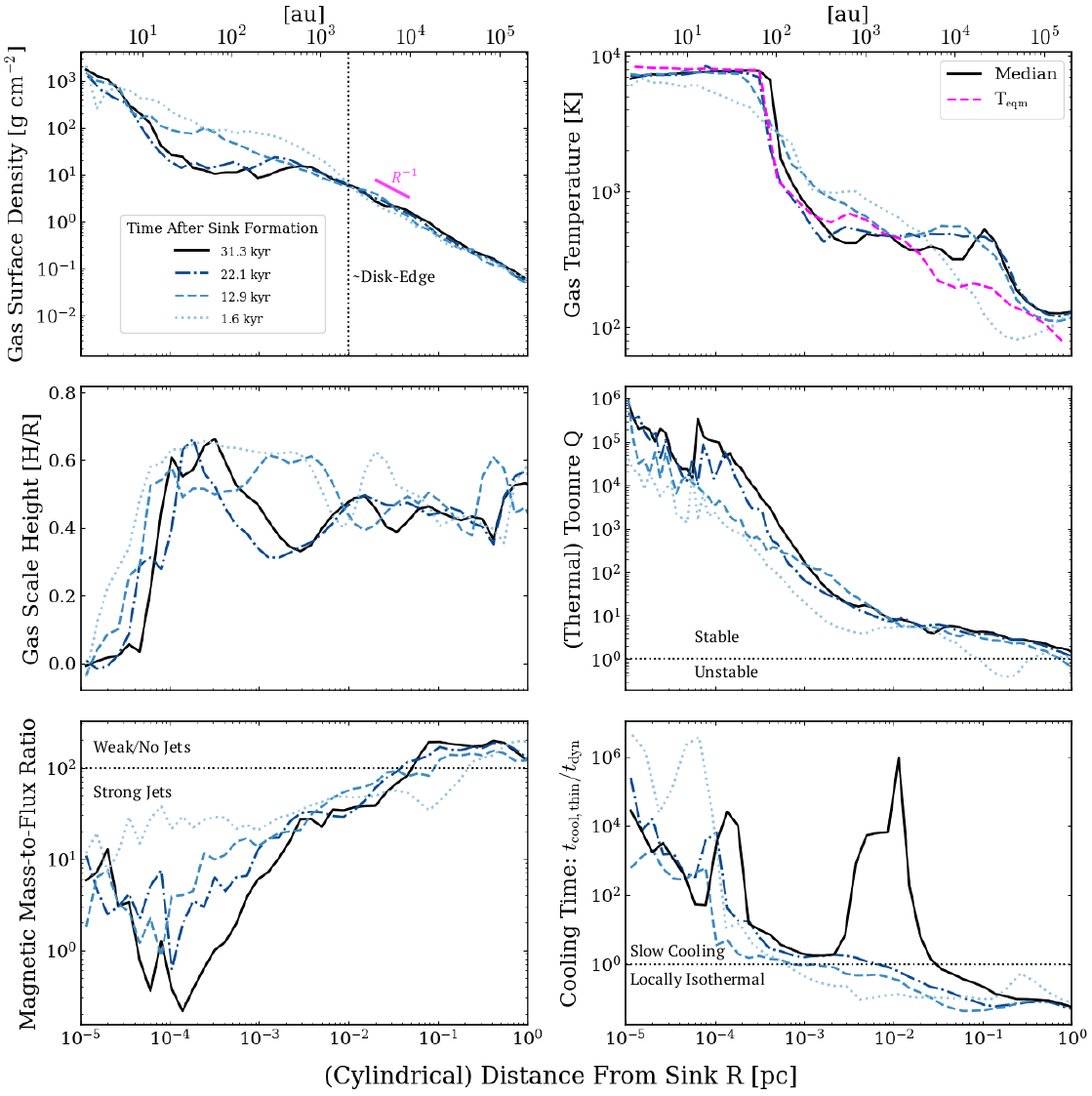}
\caption{We compute all profiles in concentric cylindrical annuli centered on the sink particle. For all panels, we average the values over the $15$ preceding snapshots at each plotted time ($\Delta t \sim 650$ years), which suppresses snapshot-to-snapshot fluctuations and avoids features that may be unique to a single snapshot. \textit{Top Left:} (Median) Gas surface density ($\Sigma_{gas}$) profile versus cylindrical distance from the sink. The protostellar disk extends out to $\sim 0.01 \; \mathrm{pc}$ (where the $v_\phi > v_r$ fraction drops), and we see that the surface density begins to taper out toward the outskirts of the disk. Between $\sim 10^{-3}-10^{1} \; \mathrm{pc}$, we find that $\Sigma_{gas} \propto R^{-1}$. \textit{Middle Left:} We measure the dimensionless scale-height $H/R$ of the gas (mass-weighted) in each annulus as the median $|z|$ after rotating to the angular momentum axis of that annulus. \textit{Bottom Left:} We compute the magnetic mass-to-flux ratio $\mu$, in each annulus as $\mu = \frac{(M/\Phi)}{(M/\Phi)_{\rm crit}}$, where $M$ is the gas mass in the annulus, $\Phi$ is the magnetic flux through the annulus, and $(M/\Phi)_{\rm crit} = 1/(2\pi \sqrt{G})$ is the critical value for collapse. Values of $\mu$ below the horizontal reference line satisfy the conditions for jet launching, whereas higher $\mu$ (above the line) suppresses jets given the mass-to-flux constraints from \citet{machida_formation_2013} (see Section \ref{subsec:Circumstellar Disk} for an explanation). \textit{Top Right:} Temperature radial profile showing the median temperature in each annulus. T$_{\mathrm{eqm}}$ is the equilibrium temperature set by the balance between accretion heating and optically thin radiative cooling. \textit{Middle Right:} We compute the Toomre Q of the gas in each annulus as $Q=\frac{c_s\,\kappa}{\pi G\,\Sigma_{\rm gas}}$; values $Q\lesssim 1$ indicate gravitational instability. \textit{Bottom Right:} Cooling-to-dynamical time ratio $t_{\rm cool}/t_{\rm dyn}$ for each annulus. Values $<1$ indicate the gas cools within a dynamical time, while $>1$ implies dynamical evolution outpaces cooling.}
\label{fig:disk}
\end{figure*}


\begin{figure}[ht] 
\includegraphics[width=\columnwidth]{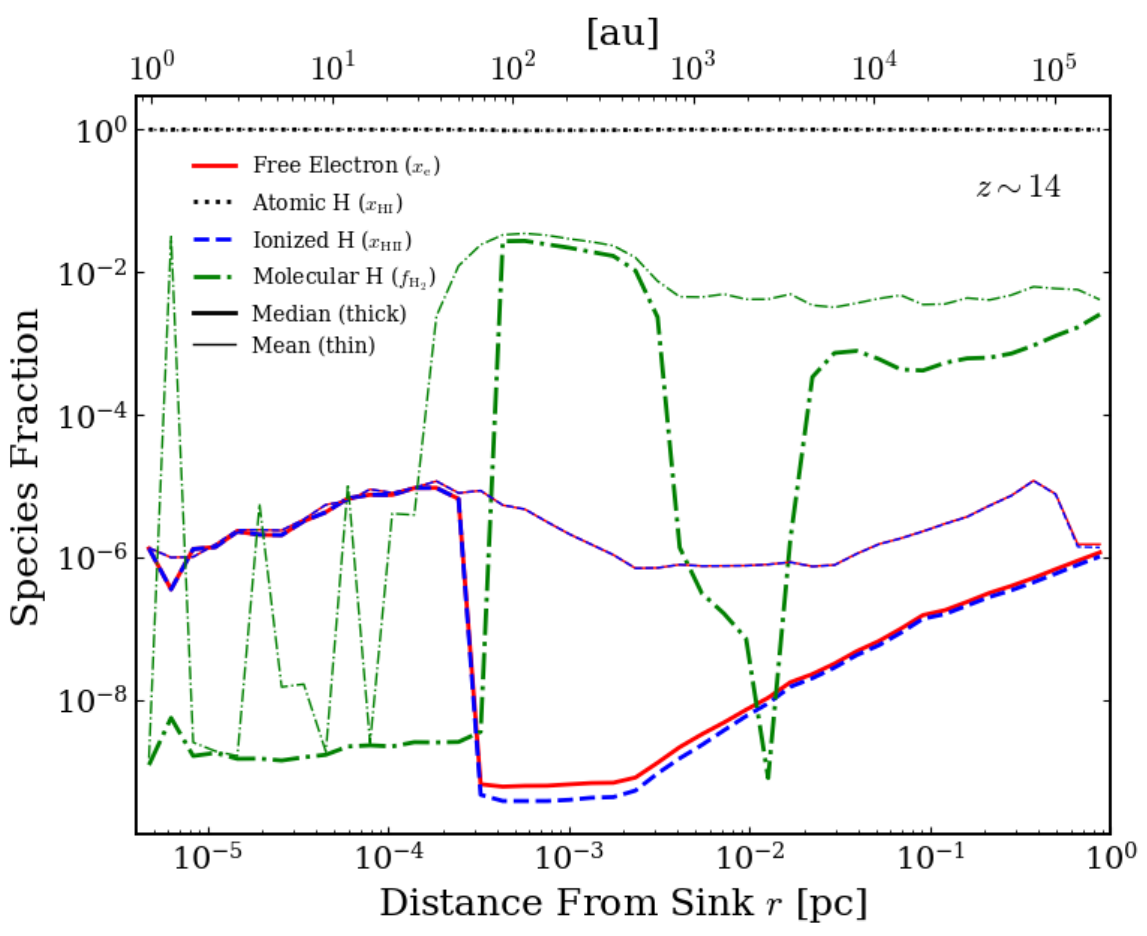}
\caption{Chemical structure of the gas surrounding the (proto)star, focusing specifically on the abundances of key species: free electrons ($x_{\rm e}$), ionized hydrogen ($x_{\rm HII}$), molecular hydrogen ($f_{\rm H_{2}}$), and atomic hydrogen ($x_{\rm HI}$). The median species fractions are shown with thick lines, while the mean values are shown with thin lines. We use the same time-averaging scheme described in the caption of Figure~\ref{fig:disk}.}
\label{fig:species fraction}
\end{figure}

The (proto)star co-forms with its circumstellar disk (Figure \ref{fig:visualization}). The (proto)star serves as the gravitational anchor of the system. As shown in the top-left panel of Figure \ref{fig:sink properties}, the (proto)star grows by accreting mass from the disk. As shown in Figure~\ref{fig:disk}, the Toomre $Q$ value in the disk,
\begin{equation}
Q = \frac{c_s\kappa}{\pi G \Sigma_{\mathrm{gas}}},
\end{equation}
where $c_s$ is the sound speed, $\Sigma_{\rm gas}$ is the gas surface density, $\kappa$ is the epicyclic frequency, and $G$ is the gravitational constant, is consistently much greater than unity at early times, indicating that the (proto)star grows from a stable disk. The disk extends to roughly $\sim 0.01\,\mathrm{pc}$ from the central star.

Figure~\ref{fig:disk} shows the radial profiles of key disk properties, including the surface density, temperature, scale height, Toomre $Q$ parameter, magnetic mass-to-flux ratio $\mu$, and the ratio of the cooling to dynamical time. All radial profiles shown in Figure \ref{fig:disk} are computed in concentric cylindrical annuli centered on the (proto)star. The magnetic mass-to-flux ratio is defined as
\begin{equation}
\label{eq:mass-to-flux}
\mu = \frac{(M/\Phi)}{(M/\Phi)_{\rm crit}},
\end{equation}
where $M$ is the gas mass in the annulus, $\Phi$ is the magnetic flux threading the annulus, and $(M/\Phi)_{\rm crit} = 1/(2\pi \sqrt{G})$ is the critical value for gravitational collapse. The cooling time is given by
\begin{equation}
t_{\mathrm{cool,thin}} = \frac{(3/2)\,nk_{B}T}{n_{\rm H}^{2}\,\Lambda(T)},
\end{equation}
where $T$ is the gas temperature, $n$ is the gas number density, $n_{\rm H}$ is the number density of hydrogen nuclei, $k_{B}$ is the Boltzmann constant, and $\Lambda(T)$ is the cooling function evaluated at the local temperature. The dynamical time is written as
\begin{equation}
t_{\mathrm{dyn}} = \sqrt{\frac{r^{3}}{G\,M_{\rm enc}}},
\end{equation}
where $M_{\rm enc}$ is the total mass enclosed within radius r. Each quantity is plotted as a function of cylindrical distance from the (proto)star, out to $1 \; \mathrm{pc}$.

From the gas kinematics, the disk extends to $ \sim 2000 \, \mathrm{au}$, where the fraction of gas particles with tangential velocities greater than their radial velocities declines significantly. Larger radii ($\gtrsim10^{4} \; \mathrm{au}$ or $\gtrsim{0.05} \; \mathrm{pc}$) represent the parent cloud and the ISM of the host.

The surface density ($\Sigma$) profile of the disk, shown in Figure \ref{fig:disk}, has two distinct regions. For radii greater than $700 \; \mathrm{au}$, the surface density is proportional to $R^{-1}$. This profile is consistent with a disk formed from slow, isothermal collapse, where a $\rho \propto r^{-2}$ outer envelope yields $\Sigma \sim \rho r \propto R^{-1}$ \citep{shu_self-similar_1977}. In contrast, within the inner regions (R $< 700 \; \mathrm{au}$), the surface density is broadly consistent with an $\alpha$-disk model \citep{shakura_black_1973}, where $\alpha$ is a dimensionless viscosity parameter. In an $\alpha$-disk, $\Sigma \propto \dot{M}\Omega/(\alpha T)$. If $\dot{M}$ and $\alpha$ vary weakly with radius and the rotation curve is Keplerian ($\Omega \propto R^{-3/2}$), a flattened $\Sigma$ profile implies that the temperature decreases sufficiently steeply with radius to offset the $\Omega$ dependence; this temperature behavior is approximately what we observe at $\sim 80 \; \rm au$ (see Figure \ref{fig:disk}). In the innermost regions of the disk, a sharp increase in surface density is observed at $\sim 20\;\mathrm{au}$ for the most recent measurement ($31.3\;\mathrm{kyr}$). Because $T$ and $\alpha$ vary only weakly in this region, the scaling $\Sigma \propto R^{-3/2}$ applies, producing a steep rise in surface density. This steep rise is not present at earlier times ($1.6 \; \mathrm{kyr}$), when the disk is not fully established and the surface density follows $\Sigma \propto R^{-1}$ at all radii.

From Figure \ref{fig:disk}, we see the disk's dimensionless scale height (H/R), which tells us about the dominant forms of vertical support at different radii. Here, $H$ (the physical scale height) is defined as half the vertical range containing the central $50\%$ of the mass in a cylindrical shell along the local rotation axis. In the innermost disk (R $< 20$ au), accretion heating raises temperatures, and H/R reflects expected, predominantly thermal support. In the intermediate regions of the disk (R $\sim 100$ au to $1000$ au), the temperature drops to between $10^2 - 10^3 \; \mathrm{K}$. In response, H/R also drops to between $\sim 0.3-0.4$ as the disk cools rapidly.

At larger radii (R $> 1000$ au), the dimensionless scale height (H/R $\sim 0.4 - 0.5$) is governed primarily by turbulent support (supersonic turbulence), and follows the scaling relation H/R$\sim c_s/v_K \sim v_t/v_K \sim 0.1(\mathrm{T}/500 \, \mathrm{K})^{1/2} \times(\mathrm{R}/100\,\mathrm{au})^{1/2}$, where $c_s$ is the sound speed, $v_t$ is the turbulent velocity, and $v_K$ is the Keplerian velocity. 

The gas temperature profile (Figure \ref{fig:disk}) shows that temperatures remain high (T $\sim 10^4$ K) in the inner disk. We compare this profile to the equilibrium temperature T$_{\mathrm{eqm}}$, which is set by the balance between accretion heating and radiative cooling ($\Gamma_{\rm heat} = \Lambda(T_{\rm eqm})$), with $\Gamma_{\rm heat}$ defined as

\begin{equation}
\Gamma_{\rm heat} = \frac{3 \dot{M}_{\rm acc, in} \Omega^2}{8 \pi n \rm H },
\end{equation}

where $\dot{M}_{\rm acc, in}$ is the inward accretion rate (to the (proto)star), 
and $\Omega$ is the orbital frequency. In this system, accretion provides the dominant heating source, while cooling occurs through optically thin radiative processes. At small radii, the accretion heating rate exceeds cooling by molecular hydrogen, whose cooling is effective only at temperatures of a few hundred kelvin. As a result, the gas heats above the H$_2$-cooled regime, leading to H$_2$ dissociation. The temperature then rises until Lyman cooling becomes effective at T$\sim 10^4 \; \rm K$, setting the plateau seen in Figure \ref{fig:disk}.
The resulting temperature profile is consistent with thermal equilibrium established by strong accretion heating and inefficient cooling. The equilibrium temperature presented should be interpreted as an order-of-magnitude estimate, since we assume the chemical abundances are in equilibrium at each snapshot and radius; nevertheless, this approach provides a reasonable estimate of the characteristic temperature profile.

Since molecular hydrogen is the primary coolant in metal-free gas, the local H$_2$ fraction regulates the ability of the disk to cool. The high temperatures in the inner disk (T $\sim 10^4$ K) suppress H$_2$ and favor ionized hydrogen over molecular hydrogen (Figure~\ref{fig:species fraction}). At $\sim 10^{-4} \; \rm pc$, both the median and mean molecular hydrogen fractions increase sharply, likely corresponding to the temperature drop at the same radius seen in Figure~\ref{fig:disk}. While both H$_2$ and HD contribute to cooling in this region, HD is the dominant coolant. 

In Figure~\ref{fig:species fraction}, we plot both the median and mean mass fractions for different species, which are not identical; the mean is dominated by regions with relatively small filling fractions. Clumpy accretion and self-shielding in the disk can create localized regions with very different species abundances, which may explain the discrepancy between the median and mean profiles. At $\sim 10^{-2} \; \rm pc$, the median molecular hydrogen mass fraction drops sharply, corresponding to the outer edge of the circumstellar disk (see Figure \ref{fig:disk}, top left). This drop is likely caused by dissociation of molecular hydrogen due to an accretion shock. The absence of a corresponding drop in the mean profile suggests that dense gas clumps are not dissociated by the accretion shock, and therefore retain high H$_2$ fractions.

\begin{figure}[b] 
\includegraphics[width=\columnwidth]{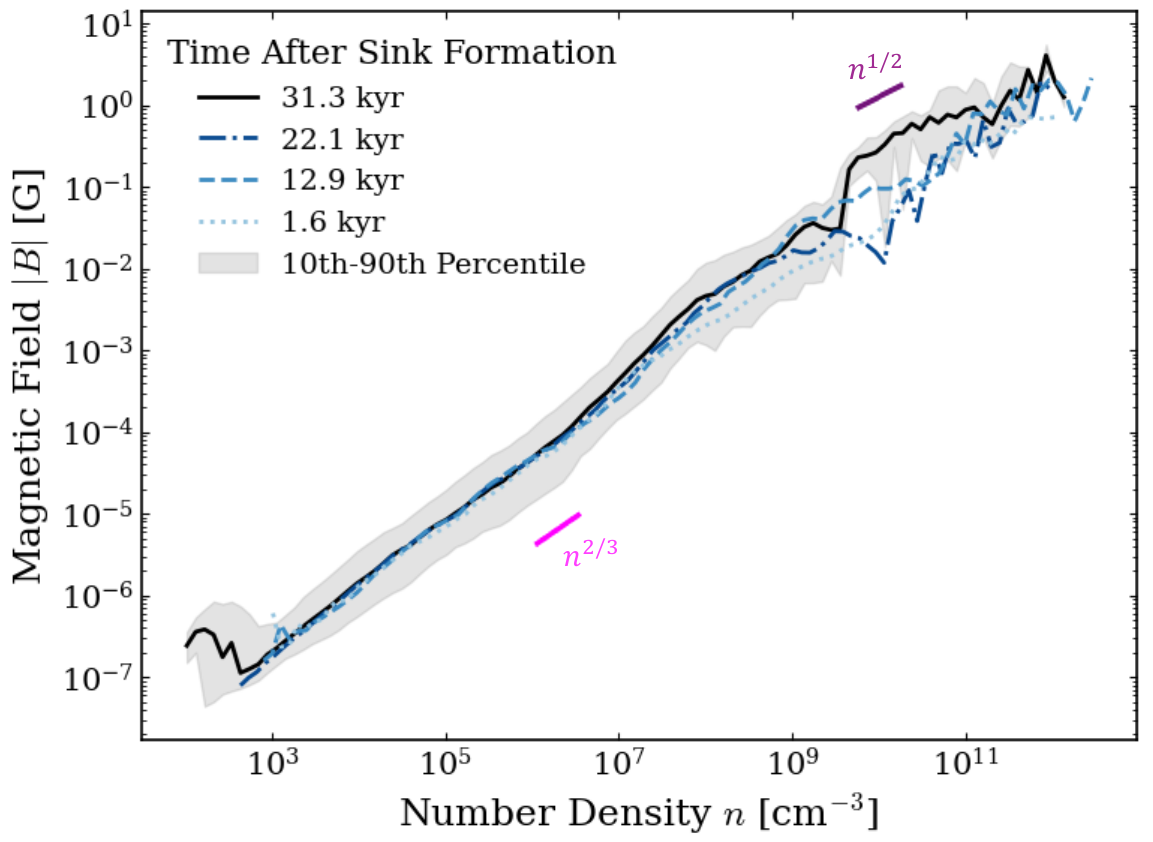}
\caption{We plot the median magnetic field strength against the gas number density. Time after sink formation is indicated in the figure. We shade the $80 \%$ inclusion interval of the magnetic field strength at $31.3 \; \rm kyr$ after sink formation. The relations $B \propto n^{2/3}$ and $B \propto n^{1/2}$ are also plotted in the figure.}
\label{fig:B vs n}
\end{figure}

The Toomre $Q$ remains well above unity ($\mathrm{Q} > 1$) across most radii in Figure \ref{fig:disk}, suggesting that the disk is gravitationally stable against large-scale fragmentation. In Figure \ref{fig:disk}, we can see that $Q>>1$ towards the inner regions of the disk, where the temperature reaches $\sim 10^{4} \, \mathrm{K}$ and thermal and rotational support are therefore the strongest. Furthermore, the ratio of the cooling time to the dynamical time ($t_{\mathrm{cool, thin}}/t_{\mathrm{dyn}}$) is above unity in the inner regions of the disk. This indicates that the gas cannot radiate away its thermal energy on a dynamical timescale, and therefore the disk remains pressure-supported rather than fragmenting. At the most recent measurement ($31.3 \; \rm kyr$ after sink formation), the sharp rise in $t_{\mathrm{cool, thin}}/t_{\mathrm{dyn}}$ coincides with a drop in the median molecular hydrogen fraction shown in Figure \ref{fig:species fraction}, consistent with a temporary loss of H$_2$ cooling.

On the other hand, we calculate the plasma $\beta$, defined as the ratio of thermal to magnetic pressure, using the thermal pressure extracted from the snapshots and the magnetic pressure
\begin{equation}
P_{\mathrm{mag}} = \frac{|B|^2}{8\pi},
\end{equation}
where $B$ is the magnetic field. Across the disk, $\beta$ lies between $10^{9}$ and $10^{10}$, while the magnetic mass-to-flux ratio, $\mu$, ranges from $10^{-1}$ to $10^{1}$. These values of $\mu$ fall within the regime expected to support jet formation in Pop.~III (proto)stars according to \citet{machida_formation_2013}. Although the trace cosmological magnetic fields used here (B$\sim 10^{-15} \; \rm G$) are weaker than those used in \citet{machida_formation_2013}, we recover similar magnetic mass-to-flux ratios in the regions where jets are present. \citet{machida_formation_2013} use idealized 3D MHD simulations of collapsing primordial clouds, in which simulation runs with mass-to-flux ratios (see Equation~\ref{eq:mass-to-flux}) in the range $\mu \sim 6 - 700$ exhibited jet launching. 

In Figure \ref{fig:B vs n} we also show the magnetic field strength $B$ as a function of the gas number density $n$. As the (proto)star grows throughout the simulation, the field approximately follows a power-law scaling $B \propto n^{\alpha}$, where $\alpha \approx 2/3$ or $1/2$. Gas that follows $B \propto n^{2/3}$ can be interpreted as undergoing isotropic spherical collapse under flux freezing \citep{mouschovias_nonhomologous_1976}. In contrast, gas that follows $B \propto n^{1/2}$ is indicative of anisotropic, flux-frozen collapse where the magnetic field is dynamically important \citep{mouschovias_nonhomologous_1976}. Therefore, these scaling relations indicate that the magnetic field is efficiently amplified during collapse and becomes dynamically important at high densities.

Our simulations neglect non-ideal MHD (NIMHD) effects as common for Pop.\ III studies \citep[e.g.,][]{sadanari_non-ideal_2023, sharda_population_2025, veenen_radiation_2025}, but we can estimate in post-processing whether this assumption is reasonable. The standard parameterization of NIMHD adds terms $\partial_{t} {\bf B} = \nabla \times [-\eta_{\rm O} \nabla \times {\bf B} - \eta_{\rm H} \nabla \times {\bf B} \times \hat{\bf B} + \eta_{\rm AD} \nabla \times {\bf B} \times \hat{\bf B} \times \hat{\bf B}] $ to the induction equation, with $\eta_{\rm O,\,H,\,A}$ defined in terms of the various chemical species abundances and collision cross sections representing Ohmic resistivity, the Hall effect, and ambipolar diffusion, respectively. We can therefore compare the characteristic timescale for $\mathcal{O}(1)$ changes to the magnetic field strength or topology via NIMHD magnetic flux transport, $t_{\rm NI} \sim r^{2}/|\eta|$, to the dynamical time $t_{\rm dyn}(r)$ at each radius, using the specific values of $\eta_{\rm O,\,H,\,A}$ calculated for the chemical abundances in the code with the definitions in \citet{wardle_magnetic_2007}. We find at the radii of interest  $t_{\rm NI}^{\rm O,\,H,\,A}/t_{\rm dyn} > (10^{10},\,10^{6},\,10^{7})$, justifying our neglect of NIMHD. This differs from metal-rich circumstellar disks, as expected, because of (1) the total lack of dust grains (a charge sink) and heavy ions, (2) the much warmer disk temperatures and lower disk densities/``puffier'' disks, and (3) much higher free electron fractions.

\begin{figure*}[t!] 
\includegraphics[width=\textwidth]{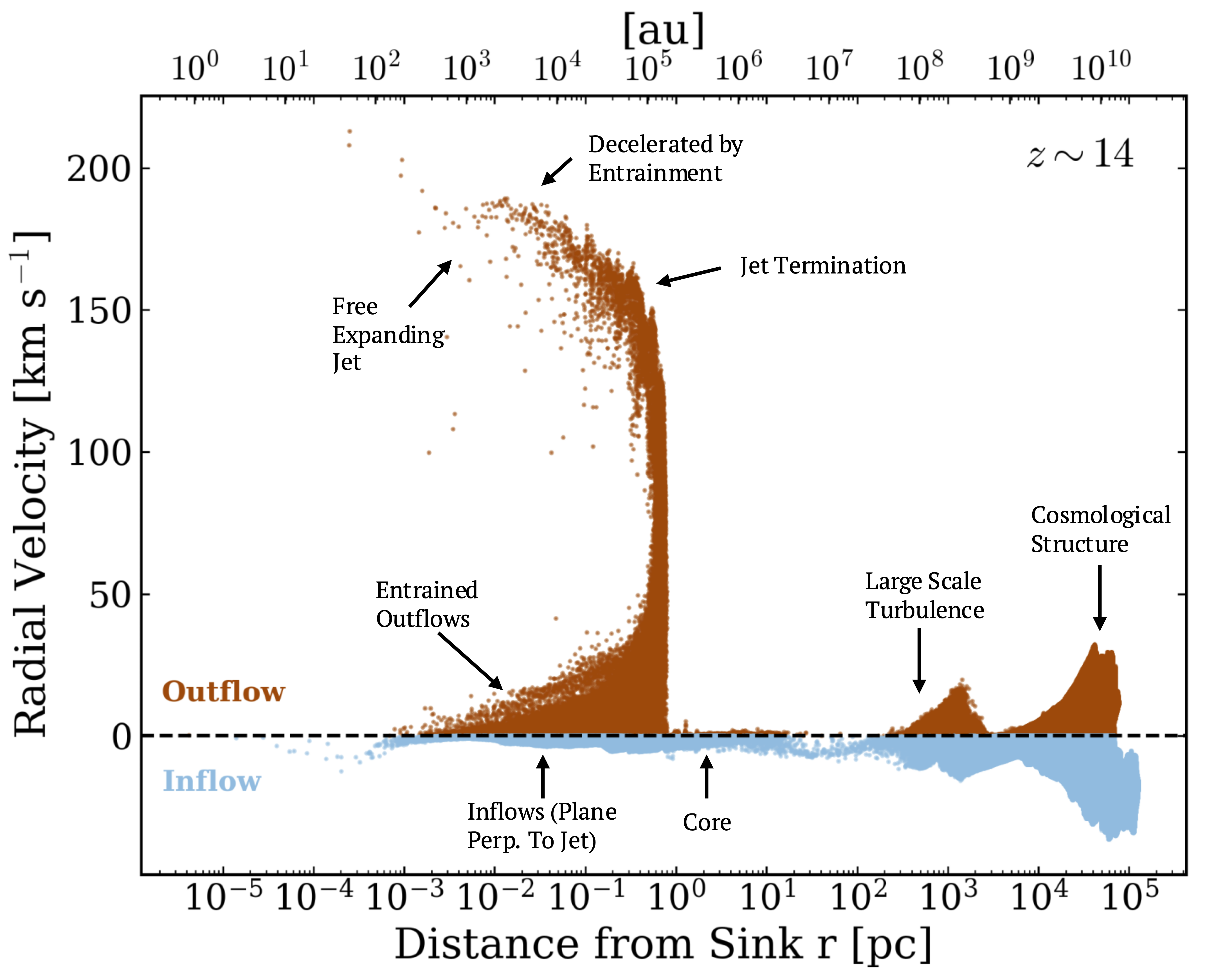}
\caption{Scatter plot showing the radial velocity of gas particles as a function of radial distance $r$ from the sink particle, at approximately $31,000 \, \mathrm{years}$ after (proto)star formation. Brown points indicate out-flowing gas (moving away from the sink), while blue points denote inflowing gas (moving toward the sink). Within the range $10^{-5}-1\, \mathrm{pc}$, particles with radial velocity greater than $50 \: \mathrm{km}\;\mathrm{s}^{-1}$ are identified as jet particles, whereas those with lower velocities correspond to disk gas. Near $1 \, \mathrm{pc}$ we observe a \enquote{termination shock} as a sharp cutoff in the radial extent of most particles, caused by the jet shock where the outflow collides with the surrounding interstellar medium.}
\label{fig:radial velocity}
\end{figure*}

\subsection{Outflow Dynamics} \label{subsec:Outflow Structure}

The (proto)star drives both collimated jets and winds that shape the surrounding gas dynamics (Figures \ref{fig:radial velocity} and \ref{fig:accretion}). About $30,000 \, \mathrm{years}$ after formation, the jets propagate into the surrounding medium and reach distances of $\sim 1 \, \mathrm{pc}$. As expected for jets in general (in simulations and observations), portions of the jet are enclosed by a cocoon-like structure (see Figure \ref{fig:visualization}), likely resulting from swept-up ambient gas and slower, energy-conserving outflow components. The disk winds emerge from the larger disk radii and carry mass and momentum into the surrounding region. 

Figure \ref{fig:accretion} shows that within the first thousand years after sink formation, the outflow accretion rate rises rapidly to $\sim 10^{-3}\, M_{\odot} \, \mathrm{yr}^{-1}$. This sharp increase corresponds to the formation of the jet, which begins expelling material from the (proto)star. The outflow rate then gradually declines to $\sim 10^{-5}\, M_{\odot} \, \mathrm{yr}^{-1}$ by $5000 \, \mathrm{years}$ after sink formation. This decline may result from the jet clearing out the nearby gas.

At the end of the simulation, the radial profile of the accretion rate shows that inflow dominates close to the protostar, while outflows become increasingly important at larger radii ($\sim 10^{-1}-10^{0} \; \rm pc$), with the largest radii ($\sim 10^{3}-10^{5} \; \rm pc$) characterized by more symmetric turbulence and cosmological structure (see Figures \ref{fig:accretion} \& \ref{fig:radial velocity}). By $\sim 10^{-2}\,\mathrm{pc}$, the inflow rate rises to $\dot{M}_\mathrm{in}\sim6\times10^{-4}\,M_{\odot}\,\mathrm{yr^{-1}}$, indicating strong accretion onto the disk. At $\sim 5 \times 10^{-3}\; \rm pc$, the outflow rate $\dot{M}_\mathrm{out}$ increases and briefly surpasses the inflow rate, suggesting that the jets and winds begin to push back against the surrounding gas. At larger distances ($r\gtrsim2\times10^{-2}\,\mathrm{pc}$), inflow again dominates as the collapse of the envelope continues.

Although $\dot{M}_{\mathrm{jet}}$ is small and scales only with $\dot{M}_{\star}$, the jets can still carry substantial momentum and therefore influence gas at much larger radii \citep{guszejnov_starforge_2021}. Protostellar jets are known to have significant effects on the IMF through this momentum-driven channel. The momentum loading, $\dot{M}_{\mathrm{jet}} v_{\mathrm{jet}}$, allows the jet to sweep up mass at a rate $\dot{M}_{\mathrm{swept}} \sim \eta \dot{M}_{\star}$, where $\eta \sim \sqrt{R_{\mathrm{jet}}/R_{\star}}$ at the radius where the jet couples effectively to the surrounding gas. In our case, this coupling occurs at $R \sim 0.5\,\mathrm{pc}$ (as seen in Figure \ref{fig:radial velocity}), giving $\eta \sim 1000$, implying that the jet greatly influences gas dynamics well beyond its launch region.

\section{Conclusions} \label{sec:Conclusions}
This paper presents the first cosmological RMHD simulation of a single Pop.~III (proto)star using the FORGE'd in FIRE framework. We follow the formation and large-scale dynamical evolution of a (proto)star from its parent halo down to au scales, resolving its circumstellar disk, jets, and winds at a gas particle resolution of $\sim 10^{-3} \, M_{\odot}$ and a length scale of $\sim10^{-4} \, \mathrm{pc}$.

The (proto)star forms at $z \sim 14$, with an initial inflow accretion rate of $\dot{M} \sim 5 \times 10^{-3} \,M_{\odot}\,\mathrm{yr^{-1}}$, which stabilizes between $\dot{M} \sim 3 \times 10^{-4}$ and $5 \times 10^{-3}\,M_{\odot}\,\mathrm{yr^{-1}}$ over $\sim 31,000 \, \mathrm{years}$. Outflows enhance the outward accretion rate 
to $\sim 10^{-3}\,M_{\odot}\,\mathrm{yr^{-1}}$ within the first thousand years, and the rate continues to fluctuate between $10^{-5}$ and $10^{-3} \,M_{\odot}\,\mathrm{yr^{-1}}$ throughout the simulation. The (proto)star reaches $\sim 27\,M_{\odot}$ and enters the Shell D–burning phase. A rotationally supported disk forms, extending to $\sim 0.01\,\mathrm{pc}$, and remains gravitationally stable throughout the simulation. Protostellar outflows appear within the first $500\,\mathrm{years}$ and interact with the surrounding infalling gas, regulating accretion onto the star.

The disk's thermal structure determines its stability. 
Local thermal equilibrium between accretion heating and radiative cooling sets the temperature in the disk. 
The disk remains gravitationally stable, with Toomre 
Q$>>1$, due to turbulent support. This turbulence, likely a form of gravitoturbulence, provides non-thermal pressure support that maintains a stable disk. 

Jets drive outflows that collide with infalling gas, forming termination shocks and reducing inward accretion by roughly an order of magnitude, while leaving the circumstellar disk stable and unfragmented. Uncertainties in the assumed seed magnetic field do not alter the jet behavior: even trace cosmological B-fields of $\sim 10^{-15} \; \mathrm{G}$ on kpc scales are sufficient to easily exceed the magnetic mass-to-flux ratio needed for strong jets.   

In subsequent papers, we will analyze a companion simulation that follows the formation and evolution of a group of Pop.~III stars from the same cosmological initial conditions. Preliminary results from that simulation reveal a stellar population with a wide mass range, from sub-solar masses to over $100 \, \mathrm{M}_{\odot}$. We find that all stars form in binary or higher-order systems. Subsequent analysis will focus on the predicted IMF and the supernovae from the most massive stars. We will track the metal-rich ejecta from these supernovae to determine their impact on the host halo and their contribution to the enrichment of the early universe.

\begin{acknowledgments}
We thank Stella S.R. Offner and Isaac Cheng for useful discussions.
Numerical calculations and plots in this paper were run on the Texas Advanced Computing Center (TACC) allocation AST21010.
Support for P.F.H. was provided by a Simons Investigator grant. 
C.-A.\ F.-G. was supported by NSF through grants AST-2108230 and AST-2307327; by NASA through grants 80NSSC22K1124, 21-ATP21-0036 and 23-ATP23-0008; by STScI through grant JWST-AR-03252.001-A; and by BSF grant \#2024262.
\end{acknowledgments}

%

\bibliographystyle{apsrev4-2}
\bibliography{popIII}{}

\end{document}